\documentclass{aastex62}

\received{}
\revised{}
\accepted{}
\submitjournal{ApJL}

\shorttitle{On the impact origin of Phobos and Deimos}
\shortauthors{Hyodo \& Genda}

\begin{document}

%Title
\title{Implantation of Martian materials in the inner solar system by a mega impact on Mars}

%Authers
\correspondingauthor{Ryuki Hyodo}
\email{hyodo@elsi.jp}
\author{Ryuki Hyodo}
\affil{Earth-Life Science Institute/Tokyo Institute of Technology, 2-12-1 Tokyo, Japan}

\author{Genda Hidenori}
\affiliation{Earth-Life Science Institute/Tokyo Institute of Technology, 2-12-1 Tokyo, Japan}



\begin{abstract}
Observations and meteorites indicate that the Martian materials are enigmatically distributed within the inner solar system. A mega impact on Mars creating a Martian hemispheric dichotomy and the Martian moons can potentially eject Martian materials. A recent work has shown that the mega-impact-induced debris is potentially captured as the Martian Trojans and implanted in the asteroid belt. However, the amount, distribution, and composition of the debris has not been studied. Here, using hydrodynamic simulations, we report that a large amount of debris ($\sim 1\%$ of Mars' mass), including Martian crust/mantle and the impactor's materials ($\sim 20:80$), are ejected by a dichotomy-forming impact, and distributed between $\sim 0.5-3.0$ astronomical units. Our result indicates that unmelted Martian mantle debris ($\sim 0.02\%$ of Mars' mass) can be the source of Martian Trojans, olivine-rich asteroids in the Hungarian region and the main asteroid belt, and some of which even hit early Earth. A mega impact can naturally implant Martian mantle materials within the inner solar system.
\end{abstract}

%% Keywords
\keywords{minor planets, asteroids: general, minor planets, asteroids: individual (A-type asteroids), meteorites, meteors, meteoroids, planets and satellites: composition, planets and satellites: formation}


\section{Introduction} \label{sec:intro}
Olivine is a major mineral of the Martian upper mantle with $\sim 60$ wt\% \citep{Ber97,Zub01}. Also, at the surface of Martian grabens, such as Nili Fossae, an olivine-rich signature is detected \citep{Hoe03,Mus09}. Interestingly, seven out of the nine known Martian Trojans, called the Eureka family, show olivine-rich spectral features \citep{Pol17}. In addition, the rare A-type asteroids orbiting within the Hungarian region ($\sim 7$\% of the total mass) and within the main asteroid belt region ($\sim 0.4$\%) have olivine-rich spectral features \citep{Dem13}. The origin of A-type asteroids is still debated even though some could be mantle materials fragmented from differentiated parent objects \citep{San14}.\\
 
More and more studies suggest that a mega impact occurs on Mars $-$ a single giant impact could simultaneously explain the formation of the Borealis basin \citep{Mar08} including its True Polar Wander \citep{Hyo17b}, spinning up Mars to current orbital period of $\sim 25$h \citep{Cra11,Hyo17b} and formation of Martian moons, Phobos and Deimos \citep[][]{Cit15, Ros16,Hes17,Hyo17a,Hyo17b}. Then, this giant impact can potentially distribute the impact-generated debris within the inner solar system similar to what happend by a giant impact on the Earth that may form the Moon \citep[e.g.][]{Bot15}.
\\
  
Recently, \cite{Pol17} suggested that the Martian Trojans may be captured from Martian mantle debris distributed near the orbits of Mars by a mega impact through an efficient capture during a sudden orbital change of Mars \citep{Sch05}. Such orbital change occurs due to the gravitational interactions with other planetesimals \citep{Bra17}. They also proposed that the rare A-type asteroids originated from the impact debris of the Martian mantle. However, it is not clear if a mega impact is indeed capable of producing a sufficient amount of ejecta with appropriate compositions and orbits that would explain the Martian Trojans and the rare A-type asteroids. Moreover, a giant impact is an energetic process that melts/vaporizes materials and thus the primordial Martian olivine-rich signature may be lost. Here we investigate the compositional and thermodynamic properties of the impact ejecta as well as their mass and heliocentric orbits, and discuss the possibility of forming Martian Trojans and rare A-type asteroids. In section \ref{sec:method}, we explain our giant impact simulations. In Section \ref{sec:impact} we present our numerical results. Finally, our conclusions and discussions are presented in Section \ref{sec:conclusion}.\\

%%%%%%%%%%%%%%
%%       Section 2
%%%%%%%%%%%%%%
\section{Numerical methods and models} \label{sec:method}
We investigate the mass, composition, thermodynamic properties and heliocentric orbits of the post-impact debris produced by the dichotomy-forming mega impact using the data obtained from smoothed particle hydrodynamics (SPH) simulations. We used the GADGET-2 code \citep{Spr05} that includes tabulated equations of state \citep{Cuk12}. M-ANEOS equation of state \citep{Mel07} is used to model a differentiated object; its core is represented by pure iron and its mantle is represented by pure forsterite. This code includes self-gravity, but does not include material strength. For a huge impact ($> 100$ km in radius), the effects of material strength are negligible \citep{Jut10,Gen17}. Note that M-ANEOS may underestimate the melting/vaporization fraction for a high pressure regime ($> 100$ GPa where impact velocity of $> 8$ km s$^{−1}$) \citep{Kra12} but, our nominal case of 6 km s$^{-1}$ impact (see below) would correctly predict the melting fraction.\\ 

Mars is modeled as a differentiated object with a core-to-mantle mass ratio of 0.3 and a total mass of $6.0 \times 10^{23}$ kg and the impactor is modeled as an undifferentiated forsterite body with a mass of $m_{\rm imp}=1.68 \times 10^{22}$ kg \citep{Mar08}. The total number of SPH particles in the simulation is $N = 3 \times 10^5$ or $3 \times 10^6$. We placed the SPH particles in a three-dimensional lattice (face-centered cubic) for Mars and the impactor. Before the impact simulations, we carried out SPH simulations for gravitational relaxation for Mars and the impactor independently, so that the two bodies achieve their hydrostatic equilibrium. Initial entropy of Mars and the impactor is set to be 2000 J K$^{-1}$ kg$^{-1}$, which corresponds to 680 K at the surface of Mars. For our canonical dichotomy-forming impact (impact energy of $E_{\rm imp}=3 \times 10^{29}$ J), we apply an impact angle of 45 degrees and an impact velocity of 1.4 times the mutual escape velocity ($v_{\rm imp} \sim 6$ km s$^{-1}$) \citep{Mar08,Hyo17a,Hyo17b} with a total of $3 \times 10^6$ SPH particles. The typical smoothing length in our SPH calculation is about $\sim 50$ km. We also carried out other impact simulations for the impact angle of 30 and 60 degrees to see the effect of the impact angle. We also examined the dependence on mass ratio and impact velocity by considering the dichotomy-forming impact energy of $E_{\rm imp}=3-6 \times 10^{29}$ J \citep{Mar08} where $m_{\rm imp}=6.0 \times 10^{21}$ kg and $v_{\rm imp}=10$ km s$^{-1}$ ($E_{\rm imp}=3 \times 10^{29}$ J) and $m_{\rm imp}=3.36 \times 10^{22}$ kg and $ v_{\rm imp}=6$ km s$^{-1}$ ($E_{\rm imp}=6 \times 10^{29}$ J) with an impact angle of 45 degrees. In these cases, we used $3 \times 10^5$ SPH particles. Impact simulations are performed in an isolated system without the gravity of the Sun. We used snapshots 20h after the impact. Note that, we confirmed that the results do not significantly change as long as we use the snapshots 5h after the impact.\\

In this paper, we are interested in the particles that are escaping from Mars’ gravity and we categorize particles that are not gravitationally bound to Mars as ejected debris particles \citep{Hyo17a,Hyo17b}. In order to calculate the heliocentric orbits (semi-major axis, eccentricity and inclination) of the ejected particles, we use positions and velocities of the ejected particles obtained from SPH simulations as input. If we know the direction of impact in the Sun-centered frame, we can uniquely define the orbits by considering the relative motions of the debris to Mars. However, the direction of impact that occurs in real life is very difficult to constrain. Giant impact is essentially isotropic \citep{Kok12} and thus we decided to assume that the debris are isotopically distributed in the radial direction in the impact plane to take into account the nature of isotropic impact direction, similar to the discussion in previous work \citep{Jac12, Bot15}. Thus, we have converted the components of the positions and velocities (components of the impact plane) of all the ejected particles into the radial component and re-distributed them in an isotropic manner to calculate semi-major axis and eccentricity of particles (Figure 4). The inclinations are directly calculated from the vertical positions and velocities from the impact plane by using the data obtained from SPH simulations (Figure 4).\\

%%%%%%%%%%%%%%
%%       Section 3
%%%%%%%%%%%%%%
\section{Debris from dichoromy-forming impact} \label{sec:impact}

\begin{figure}[ht!]
\plotone{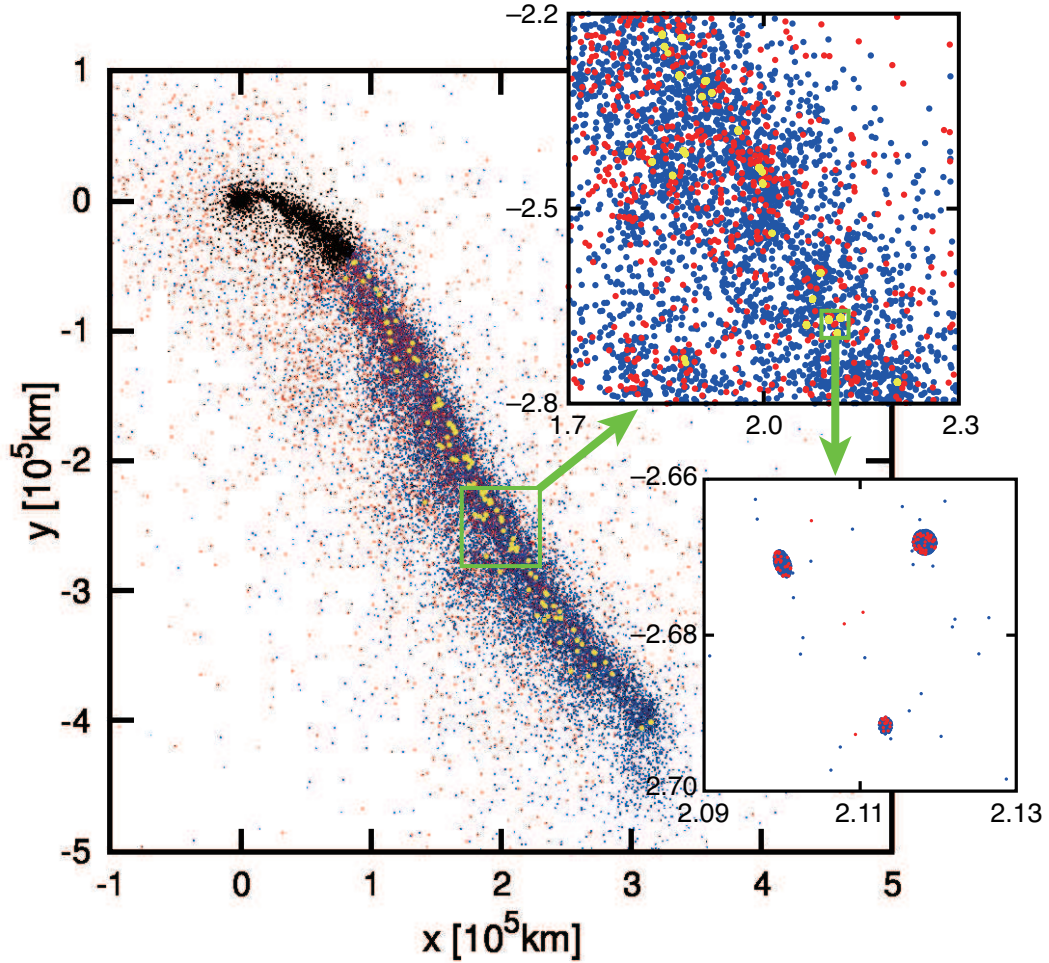}
\caption{Snapshots of SPH particles for the $45-$degree dichotomy-forming mega impact on Mars ($N=3\times10^6$ at 20h after the impact). Black points represent the components of Mars or debris disk orbiting Mars (Zoom up of the shape of Mars and the disk structures are shown in \cite{Mar08} and \cite{Hyo17a}, respectively). Blue and red points represent the component of escaping debris that are not gravitationally bound to Mars. Particles derived from impactor and Mars are colored blue and red, respectively. Yellow particles represent clumps whose number of constituent particles is larger than 10. Inserted panels show magnified views of a specific region. In the most magnified view, blue and red points are also used to represent three clumps.}
\label{surface_temp}
\end{figure}

Figure 1 shows a snapshot of the mega impact simulations ($N=3\times10^6$, impact angle of 45 degrees at 20h after the impact). A large amount of impact ejecta is produced ($\sim 42,000$ ejected particles that are not gravitationally bound by Mars) and the ejecta contains not only impactor materials, but also Martian materials. Gravitational clumps are also formed and distributed randomly in the ejecta. We systematically sort particles and calculate their orbits at 20h after the impact. We also investigate the provenance of the ejected material - either from Mars or the impactor. \\

%%%%%%%%%%%%%%
%%%%%%%%%%%%%%
\subsection{Mass, Composition and Thermodyanmics of the Debris}
\begin{figure}[ht!]
\plotone{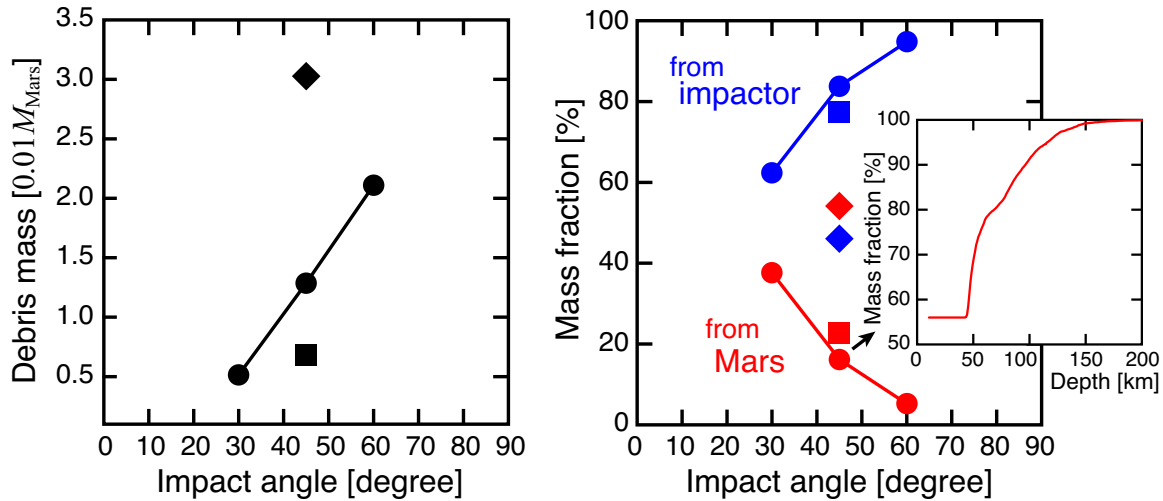}
\caption{Mass of escaping debris (left) and its composition (right) as a function of impact angle for the dichotomy-forming impact ($E_{\rm imp}=3-6 \times10^{29}$ J with $N=3 \times 10^5$ and $3 \times10^6$). Circles represent those $m_{\rm imp}=1.68 \times 10^{22}$ kg and $v_{\rm imp}=6$ km s$^{-1}$, respectively. Squares represent those $m_{\rm imp}=6.0 \times 10^{21}$ kg and $v_{\rm imp}=10$ km s$^{-1}$, respectively. Diamonds represent those $m_{\rm imp}=3.36 \times10^{22}$ kg and $v_{\rm imp}=6$ km s$^{-1}$, respectively. The red and blue color in the right panel indicates the mass fraction of Martian material and impactor material in the total debris mass, respectively. Small right panel shows cumulative fraction of the ejected particles that originated from Mars against their original depth from the surface of Mars ($N = 3 \times10^6$ with an impact angle of 45 degrees).}
\label{surface_temp}
\end{figure}

Figure 2 shows the mass of the ejected debris and their compositions in our nominal case. By changing the impact angles to a value between $30-60$ degrees, the debris mass differs between $5\times10^{-3}$ and $5\times10^{-2} M_{\rm Mars}$ and the fraction of the Martian material differs between $5-40$ wt\%. Statistically, a $\sim 45$ degree impact is the most likely \citep{Sho62}, and is the sweet spot for the Mars hemispheric dichotomy formation \citep{Mar08,Hyo17a}. For the case of 45 degree-impact high-resolution simulations ($N=3\times10^6$), debris with a mass of $\sim 10^{-2} M_{\rm Mars}$ is produced and about $\sim 20$ wt\% originates from Mars and the rest originates from the impactor. Figure 2 also shows the original location of the Martian material that is excavated by the impact. We found that about half of the Martian material originates from deep inside Mars, between 50 km and 200 km in depth (the Martian mantle). Thus, our simulations indicate that about 10 wt\% of the total debris is Martian mantle material. Note that the mass of pre-impact Mars is not significantly changed by this mega impact.\\

\begin{figure}[ht!]
  \epsscale{0.50}
\plotone{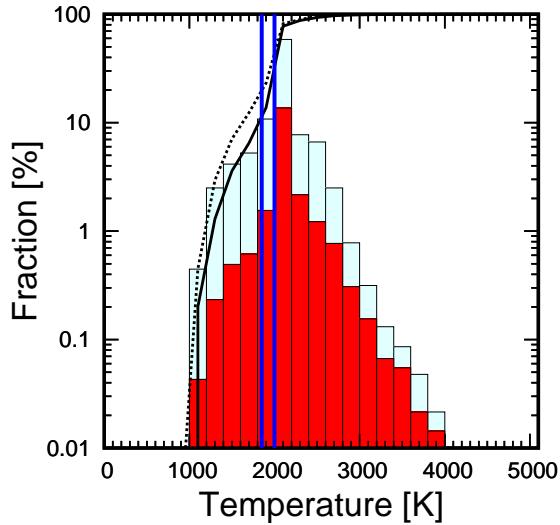}
\caption{Temperature distribution of the ejected escaping particles at 20h after the impact for the case of 45-degree impact ($N=3 \times 10^6$). The width of the bin is set to be 200 K. The cyan color represents all the ejected particles and the red color represents those of only Martian material. Dotted and solid curves represent the cumulative fraction of all the ejecta and that of Martian material. The solid blue vertical lines represent solidus (1850 K) and liquidus (2000 K) temperatures for olivine (Mg$_{\rm 0.75}$,Fe$_{\rm 0.25}$)SiO$_{4}$) .}
\label{surface_temp}
\end{figure}

Here we focus on the thermal state of ejected debris. Figure 3 shows the temperature distribution of ejected escaping debris, which ranges between 1000 K and 4000 K with a peak around 2000 K. Although this temperature distribution is similar to that of the debris disk orbiting Mars \citep{Hyo17a}, escaping debris have a somewhat wide temperature distribution because of the large variation of shock heating intensity during a mega impact. Since Mg\# (= Mg/(Mg+Fe) in mol) of bulk silicate Mars was estimated to be $\sim 75\%$ \citep{Elk03}, here we simply consider the (Mg$_{\rm 0.75}$,Fe$_{\rm 0.25}$)SiO$_{4}$ olivine solid solution as a major mineral of the Martian upper mantle. Its solidus and liquidus temperatures are about 1850 K and 2000 K \citep{Bow35}, respectively. Therefore, $\sim 10\%$ of escaping debris derived from the Martian mantle completely avoids melting, but $\sim 70\%$ of it undergoes complete melting (see Figure 3). Taking into account partial melting, we can estimate that $\sim 20\%$ of escaping debris derived from the Martian mantle avoids melting and preserves their primitive mineralogy. We also found that changing the impactor's mass and impact velocity within the possible range (see Section 2) does not significantly change this unmelted fraction. Note also that in our simulations, we have arbitrary set the initial surface temperature of Mars as $680$ K since the epoch and surface condition of Mars are not well constrained when the mega impact occurs. However, we have confirmed that this unmelted fraction is $>10$ \% when the initial surface temperature is $< 1100$ K and the fraction increases up to $\sim 30$ \% as the surface temperature decreases to $\sim 400$ K. Further investigation is required to constrain the surface condition of ancient Mars. Thus, the unmelted Martian mantle material (olivine-rich material) is estimated to be about 2\% of the total ejected mass ($\sim 1.7\times10^{20}$ kg). This mass is much larger than those of A-type asteroids found in the Hungarian region ($\sim  2.8\times10^{15}$ kg) and the main asteroid belt ($\sim 8.9\times10^{18}$ kg) \citep{Dem13}. Thus, our results imply that olivine-rich spectral features found among Martian Trojans \citep{Pol17} and the rare A-type asteroids \citep{Dem13} is attributed to this preserved olivine ejected from the Martian mantle during a mega impact.\\

%%%%%%%%%%%%%%
%%%%%%%%%%%%%%
\subsection{Orbits and Size of the Debris} 
\begin{figure}[ht!]
\plotone{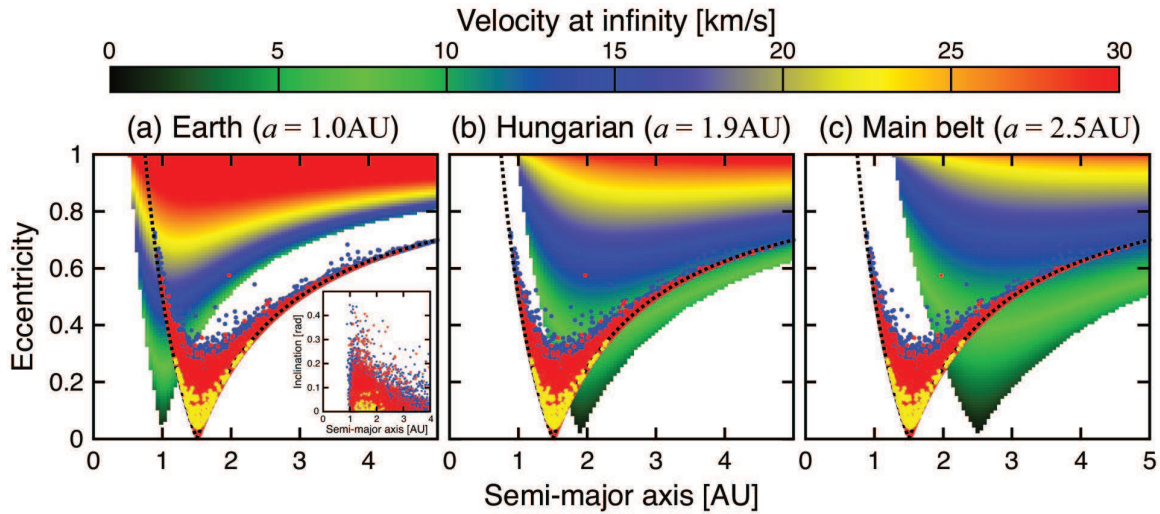}
\caption{Orbital elements of the impact debris (in the case of $N = 3 \times 10^6$ and impact angle of 45 degrees at 20h after the impact) and relative velocity between the impact debris and Earth (left panel), Hungarian (middle panel) and main belt (right panel). Red and blue particles indicate those of Martian and impactor material. Yellow particles represent unmelted Martian mantle material whose temperature is below liquidus (2000 K) for olivine ($\sim 2\%$ of the total debris). Here we assume the orbits of the debris are co-planar and circular orbits with Earth ($a=1.0$ AU), Hungarian ($a=1.9$ AU) and the main belt ($a=2.5$ AU), respectively. Two dotted lines represent orbits whose pericenter and apocenter equals $a=1.5$ AU and $e=0$. The color contours represent the relative velocity between the impact debris and Earth, Hungarian and main belt, respectively. In small panel in left panel, inclinations of the debris are shown.}
\label{surface_temp}
\end{figure}

Figure 4 shows the heliocentric orbits of the debris at 20h after the impact, assuming the impact occurs at the current location of Mars. Here, we assumed that a mega impact happened on the Martian orbital plane. The debris are widely distributed within the inner solar system between about $0.5-3.0$ astronomical units (AU) with initial large eccentricity up to around $\sim 0.6$. The inclination from the impact plane is pumped up by the impact and is distributed up to about $\sim0.3$ radian. The exact timing and radial location of Mars when the dichotomy-forming impact occurs is not well constrained. Mars may form at a greater heliocentric distance ($>2$ AU) and is then scattered inward to the current location ($\sim 1.5$ AU) \citep{Bra17}. This can explain the fact that Mars has a distinct isotopic composition from the Earth \citep{Tan14,Nug15}. If the impact occurs at a larger semi-major axis before the inward scattering, the initial distribution of the debris also develops a larger semi-major axis, but the eccentricity and inclination distribution remain essentially the same.\\

\begin{figure}[ht!]
\plotone{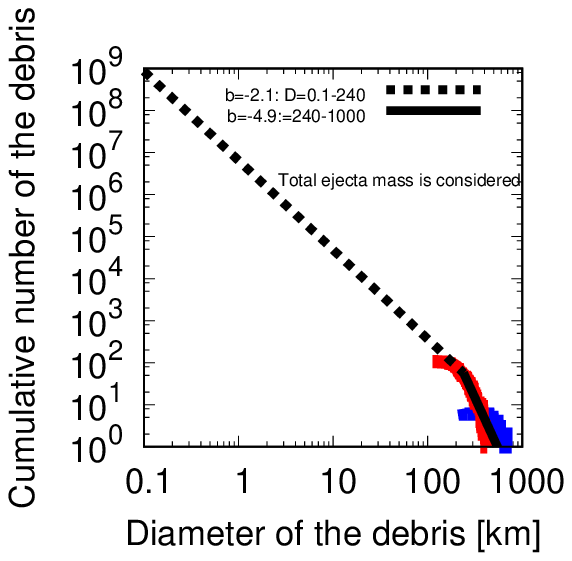} 
\caption{The initial cumulative SFD of the ejected debris in our simulations (20h after the impact in the case of impact angle of 45 degrees). The largest clump consists of 592 SPH particles. Only clumps that consists of more than 10 SPH particles are considered and are plotted by red ($N=3\times10^6$) and blue ($N=3\times10^5$) lines. The solid black line represents a cumulative power law SFD (a slope of $-4.9$) fitted between $D=240-500$ km using the data obtained from our SPH simulations (red lines). The dotted line represents that fitted between $0.1-240$ km assuming a minimum size of $D=0.1$ km and the mass balance to satisfy the total ejecta mass, which connects to the solid black line.}
\epsscale{0.50}
\label{SFD}
\end{figure}

The primordial size of the Martian Trojan Eureka is estimated to be a diameter of about $D \sim 2$ km \citep{Nug15}. The initial cumulative size frequency distribution (SFD) of large impact debris ($D > 100$ km) with the largest detected diameter of $D \sim 420$ km is directly derived from our impact simulations with sufficient resolution (Figure \ref{SFD}, see also Figure 1). We found that a steep slope of $q = 4.9$ for a power law cumulative SFD fitted between $240-500$ km (whose diameter is larger than D, $N_{\rm c}(>D) = C D^{-q}$ where $C$ is constant). For smaller fragments ($D < 100$ km) where the simulation results are not directly applicable due to the resolution limit, we apply a single power law cumulative SFD that satisfies the mass balance (total debris mass) for the rest of the debris, assuming a minimum size of $D = 0.1$ km to connect the simulation results of 240 km. By doing this, we derive a slope of $q \sim 2.1$ for this size range and predict the existence of the $\sim 10^{6}$ of $ D \sim 2$ km or greater sized debris. \cite{Pol17} estimated the required number of debris of this size to explain Martian Trojans by the capture scenario as $\sim10^{5}$ by assuming a largest fragment size of 114 km and a slope of $q = –2.85$ for the cumulative SFD. Our simulations show that the total amount of ejected material by the dichotomy-forming impact is about 50 times larger than that estimated by \cite{Pol17}. Then, 2\% of the total ejected material is unmelted Martian mantle (Figure 3) and thus our results are consistent with the requirement derived from \cite{Pol17}. Direct numerical simulations using our numerical results as input is required to understand the detailed process of capturing the debris. We will leave this matter to future work.\\

%%%%%%%%%%%%%%
%%       Section 4
%%%%%%%%%%%%%%
\section{Discussion and Conclusions} \label{sec:conclusion}
In this letter, we have shown that the dichotomy-forming impact produced a large amount of debris. The impact debris may be implanted within the asteroid belt. Figure 4 shows the post-impact distributions of eccentricity and inclination of the escaping debris, which are similar to those of the Moon-forming giant impact on Earth \citep{Jac12,Bot15}. However, the heliocentric distance where a mega impact happened on Mars ($\sim 1.5$ AU or larger) should be larger than that for the Moon-forming giant impact ($\sim 1.0$ AU). Although some of the debris is expected to re-accrete onto Mars and the other terrestrial planets \citep{Gen17}, a significant fraction of the debris can reach the asteroid belt region as a direct consequence of orbits resulting from a mega impact and/or through the successive planetary perturbations and resonances \citep{Jac12,Bot15}. Our impact simulations demonstrated that the initial orbits of the debris can easily reach the asteroid belt region and thus unmelted Martian mantle material (a maximum of $\sim 2\%$ of the total debris mass) is potentially expected to settle into stable orbits as rare A-type asteroids found in the Hungarian and main asteroid belt regions. Investigation into the long-term evolution of the debris would be required to estimate the exact capture efficiency, which strongly depends on the initial orbits of the debris.\\

We also expect that the mega-impact-induced debris hit the pre-existing asteroids, which has been discussed in the context of the Moon-forming giant impact \citep{Jac12,Bot15}. Although the amount of the debris produced by the dichotomy-forming impact is one order smaller than that produced by the canonical Moon-forming giant impact (a few percent of the Earth’s mass) \citep{Can04}, the location of a mega impact on Mars should be closer to the asteroid belt region. Therefore, we expect that the mega impact event that happened on Mars is also recorded in the asteroid belt.
The impact velocity of the debris to pre-existing asteroids in the Hungarian region and in the main belt is expected to be larger than 5 km s$^{-1}$ (Figure 4), while the nominal collision velocity between asteroids is $\sim 5$ km s$^{-1}$ \citep{Bot94}. Such a high energetic impact can be recorded as a reset of $^{40}$Ar-$^{39}$Ar age and/or impact melts \citep{Kur18}, which is also expected in the case of the Moon-forming giant impact \citep{Bot15}. Since the timing of a mega impact on Mars is likely different from that of the Moon-forming giant impact \citep[$\sim 100$ Myr after CAI condensation,][]{Tou07}, one of the multiple signatures for $^{40}$Ar-$^{39}$Ar resetting age in chondrites \citep{Bot15} would be attributed to a mega impact on Mars.\\

Our simulations demonstrated that Martian materials are widely distributed within the inner solar system as a natural consequence of a mega impact and are eventually implanted as Martian Trojans and asteroids in the Hungarian and main belt. Also, the impact ejecta can strike the pre-existing asteroid belt, and thus impact signatures that show a different epoch from those of the Moon-forming impact should be recorded on some meteorites. In addition, some of the debris should be transferred to early Earth. Since all ejected materials experienced a temperature above 1000 K in our simulations (Figure 3), this material transport does not directly support the panspermia hypothesis, in which life was born elsewhere in the universe (e.g., on Mars) and was transported to Earth. However, specific types of minerals or elements on Mars, such as borate or boron and molybdenum \citep{Ste13} - which are thought to play important roles in the origin of life \citep{Ric04}, but would be lacking on early Earth - can be delivered to early Earth through this process. Future planetary explorations and the detailed data analysis of the meteorites will test our predictions.

\acknowledgments
We thank the anonymous reviewer for comments that improved the presentation of the manuscript. This work was supported by JSPS Grants-in-Aid for JSPS Fellows (JP17J01269), MEXT KAKENHI grant (JP17H06457), and the Astrobiology Center Program of National Institutes of Natural Sciences (NINS) (AB291011).



\end{document}